# Photon Generation from Quantum Vacuum using a Josephson Metamaterial

P. Lähteenmäki[1], G.S. Paraoanu[1], J. Hassel[2], and P. J. Hakonen[1]

[1]Low Temperature Laboratory, Aalto University, PO Box 15100, FI-00076 AALTO, Finland
[2] VTT Technical Research Centre of Finland, PO BOX 1000, FI-02044 VTT, Finland

**Abstract**- When one of the parameters in the Euler-Lagrange equations of motion of a system is modulated, particles can be generated out of the quantum vacuum. This phenomenon is known as the dynamical Casimir effect, and it was recently realized experimentally in systems of superconducting circuits, for example by using modulated resonators made of coplanar waveguides, or arrays of superconducting quantum intereference devices (SQUIDs) forming a Josephson metamaterial. In this paper, we consider a simple electrical circuit model for dynamical Casimir effects, consisting of an LC resonator, with the inductor modulated externally at 10.8 GHz and with the resonant frequency tunable over a range of $\pm 400$MHz around 5.4 GHz. The circuit is analyzed classically using a circuit simulator (APLAC). We demonstrate that if an additional source of classical noise couples to the resonator (on top of the quantum vacuum), for example *via* dissipative "internal modes", then the resulting spectrum of the photons in the cavity will present two strongly asymmetric branches. However, according to the theory of the dynamical Casimir effect, these branches should be symmetric, a prediction which is confirmed by our experimental data. The simulation presented here therefore shows that the origin of the photons generated in our experiment with Josephson metamaterials is the quantum vacuum, and not a spurious classical noise source.

## 1. INTRODUCTION

The picture about vacuum that emerges from modern quantum field theory is very different from that offered by classical physics. In classical field theory, the vacuum is the zero-energy state of the field, defined by the absence of any excitation. In quantum field theory, the vacuum state has a finite zero-point energy associated with it, and the uncertainty principle indicates that fluctuations exist even in this state. Due to the existence of fluctuations, the quantum vacuum can become unstable under certain perturbations, and the energy of the perturbation is converted into creation of real particles [1]. In the Schwinger effect for example, a static intense electric field can create pairs of electrons and positrons. In very strong gravitational fields, at the event horizon of black holes, the vacuum becomes unstable and energy is radiated away (Hawking effect). By the equivalence principle, an accelerated observer in the Minkowski vacuum will detect a finite-temperature field (Unruh effect).

In a previous paper, we reported the observation of the dynamical Casimir effect using a flux-biased Josephson metamaterial embedded in a microwave cavity at 5.4 GHz [2]. A non-adiabatic change in the index of refraction of the cavity (or in the electrical length) was realized by modulating the flux at values close to double the resonant frequency of the cavity. We measured also the frequency-correlated photons emitted from the cavity (which is kept at a temperature $T = 50$ mK), and we obtained a power spectrum displaying a bimodal, "sparrow-tail" distribution, with two symmetric branches. Furthermore, we demonstrated that the generated photons are squeezed below the vacuum level and that the state of the two-photon field is nonseparable. The experimental results are in excellent agreement with the theoretical predictions. An alternative realization of the dynamical Casimir effect consists in modulating an effective boundary condition, realized experimentally as a SQUID structure placed at the end of a coplanar waveguide resonator [3]. Remarkably, the experiments presented in Refs. [2] and [3] are the only existing measurements of the dynamical Casimir effect, which was predicted theoretically in 1970 [4].

## 2. MODEL FOR NOISE ANALYSIS

A SQUID array was used in Ref. [2] to test the dynamical Casimir effect. This array was coupled to the external circuit by a capacitor. The entire system behaved as an electromagnetic cavity with tunable index of refraction. A simple analog model for this system, realized with electronic components, can be constructed as an $LC$ circuit with the resonant frequency $1/2\pi\sqrt{LC}$ tunable in a range of $\pm 400$ MHz around 5.4 GHz. The inductance $L$ is modulated at a frequency of at 10.8 GHz, and classical white noise is fed in the circuit (see Fig. 1). The purpose of this simulation is to see the effect produced by spurious sources of noise on top of the quantum vacuum. In the experiment presented in Ref. [2] (see also its Supplementary Material) this could be for example additional thermal noise coming through the electromagnetic modes that couple to the cavity. Some of these modes are not directly accessible to the experimentalist (they are the so-called "internal modes") and they might not thermalize well with the rest of the sample, at the base temperature of the refrigerator ($T$ = 50 mK, corresponding to a negligible thermal occupation of 0.0056 at the frequency of 5.4 GHz).

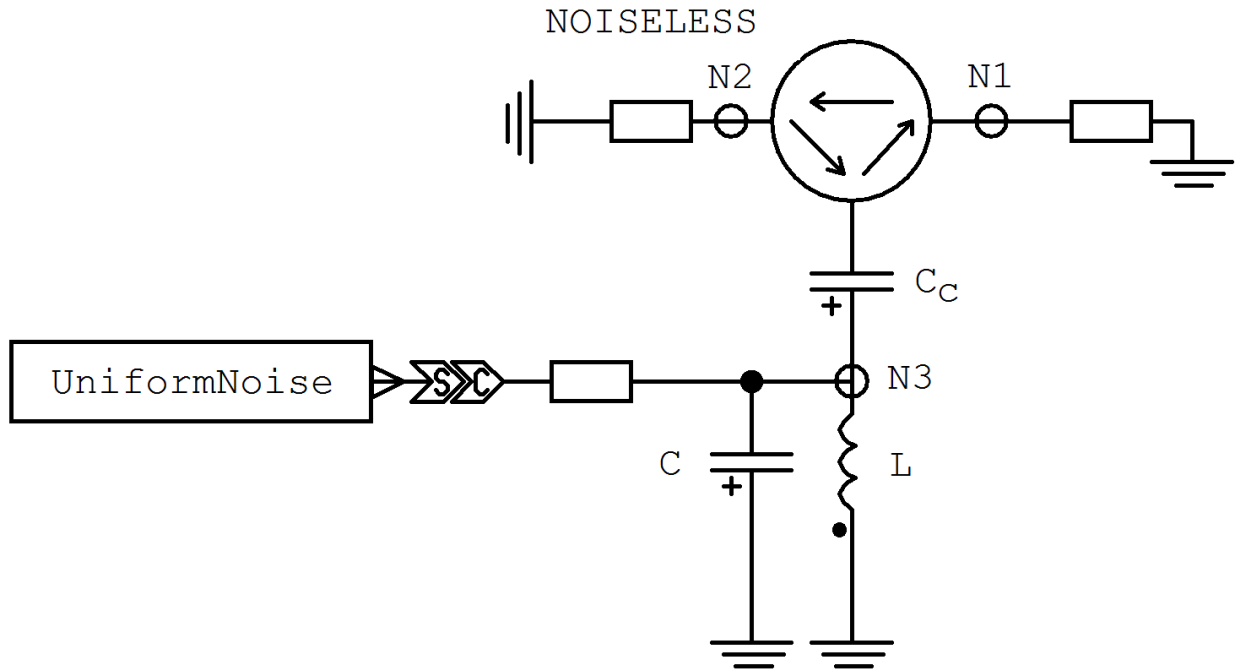

Figure 1: Schematic of the circuit used in the APLAC simulation. An $LC$ circuit is formed between node N3 and the ground. This circuit is coupled through a capacitor $C_c$ to an external circuit consisting of a circulator and two resistors. White (uniform) noise is injected in the circuit through a resistor, and the spectrum at node N1 is recorded.

The circuit is simulated using APLAC, and the spectrum at node N1 can be obtained (Fig. 2). The vertical axis $\nu$ (MHz) is the frequency at a given measurement point, while the horizontal axis $\Delta$ (MHz) is the frequency of the resonator; both frequencies are measured with respect to half the pumping frequency (10.8 GHz/2 = 5.4 GHz). The value of the capacitance used in this simulation is $C$ = 40 fF, and the coupling capacitor is $C_c$ = 5 fF. The spectrum in Fig. 2 (left) is obtained by modulating the inductor at a constant amplitude of 2 nH and at a constant frequency of 10.8 GHz around its average value. This average value is then slowly decreased from $L$ = 25.3 nH to $L$ = 21.71 nH, resulting in an increase in the $LC$ resonant frequency from 5 to 5.4 GHz.

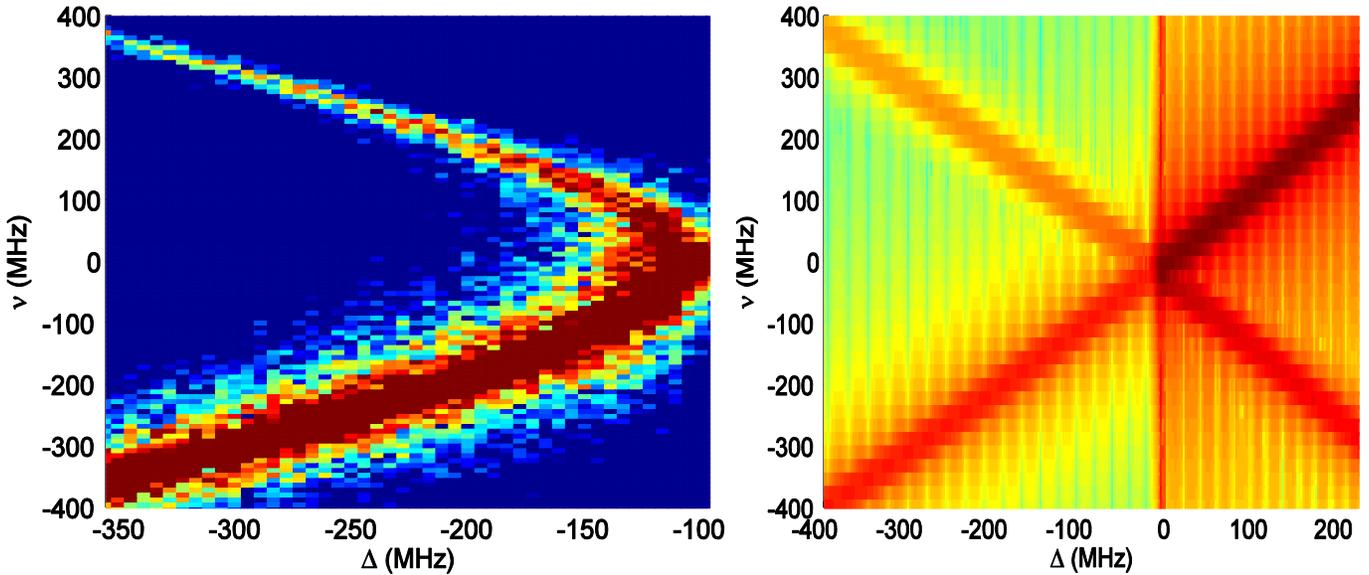

Figure 2: (left) Energy spectrum measured at the node N1 of the modulated LC circuit with a classical noise input. The vertical axis $\nu$ (MHz) is the frequency measured from half the pumping frequency (10.8 GHz/2 = 5.4 GHz), while the horizontal axis $\Delta$ (MHz) is the detuning of the resonator frequency with respect to half the pumping frequency (10.8 GHz/2 = 5.4 GHz). The figure is obtained by averaging over 100 realizations of the simulation. (right) Alternate realization of the simulation: the energy in an initially-excited and pumped LC resonator is partly kept at the resonant frequency and partly upconverted, resulting again in the appearance of two asymmetric branches in the spectrum.

An alternate realization of this simulation consists of using the same $LC$ circuit as in Fig. 1, again with the inductance being modulated at a frequency nearly twice the resonant frequency of the LC resonator, but with the capacitor $C_c$ decoupled. We start the simulation by depositing an initial amount of energy in the $LC$ resonator and we monitor the spectrum at the node N3 as the frequency of the resonator is slowly swept, reaching and passing the parametric instability threshold ($\Delta$ = 0). The resulting spectrum shown in Fig. 3 (right) looks similar to the previous simulation.

## 4. CONCLUSIONS

The energy and the fluctuations of the electromagnetic vacuum confined in a cavity have real, measurable effects: they produce an attractive force between the walls of the cavity (static Casimir effect), and, if the boundary conditions or the index of refraction are changed, photons are created (dynamical Casimir effect). The experimental demonstration of the dynamical Casimir effect by the modulation of the effective electrical length (or the index of refraction) of a cavity constitutes an important milestone in quantum field theory. To establish this effect, we should check that the input state of any mode that couples to the cavity is the quantum vacuum state. In this paper, we show by a simple circuit simulation that any spurious (uncontrolled) classical source of noise that enter the cavity (the rest of the circuit being kept at the base temperature of the dilution refrigerator) should produce a spectral pattern with strongly asymmetric branches. Since the spectrum predicted by the theory of the dynamical Casimir effect is symmetric, the result presented here shows that the observed effect cannot be explained simply by the parametric amplification of a classical signal present at the input of the device, thus supporting the quantum origin of the phenomenon. Our experiment demonstrates the potential of superconducting quantum circuits to serve as a platform for simulating effects from cosmology and quantum field theory [5].

## ACKNOWLEDGEMENT

We wish to acknowledge the Academy of Finland for financial support through projects nos. 135135 and 263457, and through the Center of Excellence in Low Temperature Quantum Phenomena and Devices (project 250280).

## REFERENCES

1. Nation, P. D., J. R. Johansson, M. P. Blencowe, and F. Nori, "Stimulating Uncertainty: Amplifying the Quantum Vacuum with Superconducting Circuits", *Rev. Mod. Phys.* Vol. 84, 1-24 (2012).
2. Lähteenmäki, P., G. S. Paraoanu, J. Hassel, and P. J. Hakonen, "Dynamical Casimir effect in a Josephson metamaterial", arXiv:1111.5608 (2011).
3. Wilson, C. M., G. Johansson, A. Pourkabirian, M. Simoen, J. R. Johansson, T. Duty, F. Nori, and P. Delsing, "Observation of the dynamical Casimir effect in a superconducting circuit", *Nature* Vol. 479, 376-379 (2011).
4. Moore, G. T., "Quantum theory of the electromagnetic field in a variable-length one-dimensional cavity", *J. Math. Phys.* Vol. 11, 2679-2691 (1970).
5. Fulling, S. A., "Aspects of quantum field theory in curved space-time", Cambridge Univ. Press, Cambridge, 1989.